\documentclass{snowmasspaper}
\begin{document}

%==============================================================================
% title page for few authors

\begin{titlepage}

% select one of the following and type in the proper number:
\date{26 Jun 2013}

%\title{CMS Paper Template - LaTeX version}
\title{A Vision on the Status and Evolution of HEP Physics Software Tools}

\begin{Authlist}
P. Canal, D. Elvira, R. Hatcher, S.Y. Jun, S. Mrenna
\Instfoot{ieph}{Fermilab, Batavia, IL 60510, USA}
\end{Authlist}

\end{titlepage}
%==============================================================================
\setcounter{page}{2}%JPP

\section{Introduction}

This paper represents the vision of the members of the Fermilab Scientific 
Computing Division's Computational Physics Department (SCD-CPD) on 
the status and the
evolution of various HEP software tools such as the Geant4 detector simulation
toolkit, the Pythia and GENIE physics generators, and the ROOT data analysis 
framework.

We do not intend to present here a complete and balanced overview of
current development, planned or potential R$\&$D activities associated with the
HEP software tools, since we would need to get many individuals and 
institutions
around the world involved in the process for the summary to be representative
of the thinking of the whole HEP community. Instead, we present a vision that
expresses opinions and interests of the members of the SCD-CPD at Fermilab.

The goal of this paper is therefore to contribute ideas to the Snowmass 2013
process toward the composition of a unified document on the current status 
and potential evolution of the physics software tools which are essential
to HEP.

%------------------------------------------------------------------------------
% Stephan Merena and Robert Hatcher
%------------------------------------------------------------------------------
\section{Generator}

\subsection{Introduction}

Monte Carlo event generators have long been an essential tool in high-energy
physics.   They are applied at almost every stage of an experimental program:
for design, for collecting data, and for interpreting data.
An example of their impact is the fact that the manual for the {\sc pythia}\ event 
generator was the most highly-cited publication in high-energy physics for 2012.

A primary goal of particle physics is to determine the Lagrangian which describes
particle interactions at short distances.   
%Measurements invariably involve long-distance physics.   
%The event generator is a tool to translate between short-distance and long-distance physics.
Since the effects of long-distance physics cannot be calculated from first principles,
the event generator must invoke models and parametrization in the translation
from short- to long-distance physics. %is translation.   
%As a concrete example, consider the measurement of the top quark mass.
%A precise measurement
%of the top quark mass and the $W$ boson mass can test whether the Standard Model is adequate to
%explain the generation of mass.  
%The $W$ boson is a color-singlet, and can be measured at colliders
%with a relatively small ambiguity.   On the other hand, the top quark is a colored particle.
%Even though the top quark decays before long-distance physics comes into play,
%no physical observable has been constructed at a hadron collider that is insensitive to the 
%neutralization of the top quark's color.   
%Event generators are thus important for bounding the effects of
%long-distance physics and determine the main systematic uncertainty on the inferred top quark mass.

In developing an experimental program, event generators are often important for determining 
what kind of accelerator (beam types and energy) is necessary for answering a particular
theoretical question.   Enough was known about the $Z$ boson to determine that a high-energy
hadron collider was the best machine for discovery and a lepton collider tuned to the measured
mass was the best machine for measuring its detailed properties.  However, the case for how
to understand electro-weak symmetry breaking is not as clear.   The fact that many viable proposals
for physics beyond the Standard Model involve weakly-interacting particles further complicates
the issue.   Event generators allow physicists to run toy experiments and make judgments on the
most promising experimental program.

Event generators also play a part in detector design, especially
when extrapolations must be made to unexplored energies.
Educated guesses must be made regarding the production of charged
and neutral energy in time and space.  For example, the 
occupancy of a silicon detector must be known to determine
an optimal distance from the beam pipe.

When running an experimental program, % at a hadron collider, 
decisions must be made on how to collect data.   Event generators
give estimates of the production of event topologies that
can be used as trigger signatures and evaluation of
alternative schemes.
%high-transverse momentum charged tracks and mini-jets.

Most detectors %at high-energy colliders 
have limited angular and energy coverage.
Particles can be lost because they are produced 
%in a forward region (near the beampipe)
particular directions,
or at too low of an energy or because of detector inefficiencies.
Event generators are used to estimate these acceptance effects.
%to translate a number of events in the active region
%of a detector into a production cross section.

\subsection{Software}
 
Event generators are integrated into the computational
framework of an experiment so that their predictions can
be treated on an equal footing with data.   This means
that the software should act as a library.  For some given
configuration data, which specify the parameters of the
physics model, the event generator returns a list of
particles (including information about the particle type,
momentum, production vertex, etc.) that can be used as an
input for a detector simulation.

In the past, event generators were mainly written in {\sc fortran}.
Currently, C++ versions of the major event generators exist,
though they are used interchangeably with some of the older
{\sc fortran} codes.   The main reason for this is that the {\sc fortran}
codes have been %are 
validated and tuned to data to provide reliable
predictions.   Thus, despite the computational advantages of
using programs written in a modern language, the physics content
of the program is an important consideration.

%Event generators are generally designed 
%to fit
%%such that they fit easily
%into the simulation chain of an experiment.  
% [RWH] duplicative of first sentence of first paragraph of this section
The codes produce
theoretical (or Monte Carlo truth) event records of the underlying
interactions, a combinations of the incoming particles, intermediate
states and the long lived final state particles leaving the interaction.
%for the bulk of
%the Standard Model processes. % on a relatively short time scale.
Usually a %, a small (but fixed) 
number of events are generated and then propagated
through a detector simulation.    
The output of the detector simulation
is a record that is mostly indistinguishable from the recorded data.
The event generator can operate in such a fashion because many of 
the physics processes can be simply coded and sampled using Monte
Carlo techniques.   However, this structure has become less 
attractive as more advanced and detailed calculations are required.
%% [RWH] what is the implied intent of this last sentence?  what is it hinting at?

In some instances it is sufficient
to generate lots of events and use more crude smearing, rather than
detailed detector simulations, to explore the potential for
discovery in various physics scenarios.  In such cases it is 
important that the generator be efficient.
In some experimental setups, such as neutrino experiments, the
interaction region is diffuse and not a central crossing point;
in such cases the generator is needed to determine where the neutrino 
probe interacts with the target nucleon within the detector apparatus. 

Most of the recent advancements in predicting Standard Model processes
has come at the stage of the event development where fixed-order, perturbation
theory can be applied.    This piece is know as the hard interaction, and the
predictions coded into the event generators are typically those at the lowest
order of perturbation theory.  The higher-order, more-accurate predictions are
computationally intensive and often require specialized code.
The other pieces of event generation, such as
the parton shower, multi-parton interactions, hadronization, 
nuclear breakup and decays must also %still
be handled by the event generator models. 

For generators used by collider experiments much work has gone into interfacing
the calculation of the hard interaction beyond the lowest order with 
the event generators.
Mostly, this is accomplished by passing the output of the
the higher-order calculation into the event generator through files.
This adds the complication that the hard-interaction files must be
coordinated with the running of the event generator.

Alternatively, software can be rewritten to perform both the calculation of 
the hard interaction and the other aspects of event generation.   This
has certain practical limitations.   One is that executable must increase in
size as the calculations become more extensive.  This can become an issue
on computing platforms with limited memory.  A second is that many of the codes
to calculate hard interactions are not engineered to fit easily into this model.

Examples of event generators used by HEP physics programs are {\sc pythia}\ and 
{\sc genie}.  The {\sc pythia}\ code is widely used in the collider programs
both past and present and is continually being updated to supply users with
desired simulations of numerous extensions to the physics models.  
The {\sc genie}\ code was developed to consolidate a number of competing neutrino 
event generators and provide a framework for adding extensions and modeling of 
more complex or rare processes.

\subsection{Future}

As more detailed calculations are required to better understand the theoretical
implications of HEP data, %collider data,
the event generators will face the problem of
complexity and timing.   Timing is usually not considered an important issue,
because detector simulations are usually orders-of-magnitude slower.   However,
there are several reasons why the execution speed of the event generators matter.
First, code can be developed and validated in proportion to execution speed.
Second, the tuning process -- when the parameters of the event generator are
varied to match data -- is facilitated by speed.
Thirdly, the calculation of systematic uncertainties for an analysis typically 
requires running the event generators several times, each with different configurations.

The problem of complexity may be solved by optimizing code design.
Currently, the theorists who develop code do not coordinate with 
computing or experimental experts.   We should promote communication between
these different communities. 

Improvements in timing can arise from the intelligent application of 
multi-core processors or GPU's.  Already, GPU's have been successfully employed in calculating
many-parton Feynman diagrams, and we should look for other, practical uses.
The possibility of performing multi-core calculations does not currently look
fruitful, because the event generator codes are structured serially -- typically
one step of a calculation must be completed before the next.   However, this
could be the result of the code design.  A successful merging of the theory
and computing communities may find avenues for improving the design of event
generator codes.

%------------------------------------------------------------------------------
% Philippe Canal
%------------------------------------------------------------------------------
\section{Root}

\subsection{Introduction}

ROOT~\cite{ref:ROOT} is an object-oriented C++ framework that is
designed to store huge amounts of data in an efficient way optimized
for very fast data analysis.  Any instance of a C++ class can be
stored, in a machine independent compressed binary format, into a ROOT
file. The ROOT TTree object container is optimized for statistical
data analysis over very large data sets by using vertical data storage
techniques. The TTree containers can span a large number of
files. These files may be on local disks, on the web or on a number of
different shared file systems. The users have access to a very wide
set of mathematical and statistical functions, including linear
algebra, numerical integration and minimization.  For example the
RooFit library allows the user to choose the desired minimization
engine.  In addition, the RooStat library provides abstractions for
most used statistical entities, supporting all their operations and
manipulations. ROOT also provides 1D, 2D and 3D histograms that
support any kind of operation and can be displayed and modified in
real-time using either a interactive C++ interpreter, other
interactive language like python or a graphical user interface.  The
final result can be saved in high-quality graphical formats like
PostScript and PDF or in bitmap formats like JPG that are easier to
include into a web page. The result can also be stored into ROOT
macros that can be used to fully recreate and rework the
graphics. Users typically create their analysis macros step by step,
making use of the interactive C++ interpreter, by running over small
data samples. Once the development is finished, they can then run
these macros, at full compiled speed, over large data sets by either
using ACLiC (the automatic compiler interface) to build a shared
library that is then automatically loaded and executed, or by creating
a stand-alone batch program. Finally, if processing farms are
available, the user can make full use of the inherent event
parallelism by running their macros using PROOF, that will take care
of distributing the analysis over all available CPUs and disks in a
transparent way.

ROOT is at the heart of almost all HEP experiments world wide. From
simulation to data acquisition, from event processing to data
analysis, from detector monitoring to event displays, it provides
essential components and building block for the experiments to
assemble their framework.  In particular all the experiment's data is
stored in ROOT files.  At the time of writing there were already more
than 177 PBs of data for just the Large Hadron Colliders experiments.
The use of ROOT extend beyond HEP to many sciences, including nuclear
physics, biology, astronomy and even in the private sector where many
finance firms use ROOT in their own data analysis and market trend
prediction software tools.

\subsection{Root for the Future}

Building on years of success, ROOT is in the mist of a significant
migration to future proof its architecture and code base.  ROOT is
replacing its existing home-grown C++ interpreter
(CINT)~\cite{ref:CINT}.  CINT has been since its inception in the
early 1990s, the premier solution for C++ interactivity, offering
access to most of the features of the C++ 2003 standard while providing a
very simple and portable to use solution to access any compiled code
from the interpreter session.  Upgrading CINT to support the large
number of extensions in the 2011 C++ standard would have required a
very large efforts.  In addition CINT architecture was not designed to
support heavily multi-tasking processing.

To address CINT's limitations, ROOT is leveraging the rise of a very
flexible and very programmable compiler framework,
LLVM~\cite{ref:LLVM}, which includes a fully C++ 2011 standard
compliant compiler (clang).  Clang and LLVM have become industry
standard tools supported and relied upon by a large set of commercial
companies including Apple, Google, Nvidia, etc.  Using clang and its
versatile libraries, ROOT is developing a new C++ interpreter (cling)
to replace CINT.

Migrating to the new interpreter will open up a host of new
opportunities.  Since it is based on a flexible and expandable
compiler framework, cling can easily be extended to support additional
languages, including objective C or NVidia's CUDA programming
language.  Rather than a classical interpreter, Cling's implementation
is closer to an interactive compiler.  Once processed, any code input
on the command line is slightly transformed and then compiled in
memory using the Clang compiler, the same compiler currently used to
build all of MacOS' components.  The seamless availability of
just-in-time compilation of C++ code enables a large set of possible
optimizations, including boosting I/O performance by optimizing
specific uses cases of the currently running application, improving
the minimization's algorithm's performance by customizing and
recompiling the hot-spots of the current search.  In addition, the
migration will enable the library and the users to start using the
many performance and code improvements allowed by the new 2011 C++
standard.

In conjunction with this major migration, ROOT is also preparing for
the newest set of hardware platforms.  Several efforts are in place to
leverage general graphical processor units, in particular to speed up
minimization algorithms.  The internal mathematical libraries are
being upgraded to take advantage of vectorized processing units,
including incorporating and using the VC (Vector Classes)
library~\cite{ref:VC}.  ROOT has been recently ported to both the ARM
and Xeon Phi architectures which should both benefit from the on-going
vectorization efforts.

As the number of individual cores offered in each processing unit is
increasing rapidly, the need for multi-threading and multi-processing
solutions is becoming greater.  ROOT is planning on offering multiple
alternative for parallelizing the Input and Output.  One medium term
alternative is to provide for fast merging of files produced in
parallel.  In this scenario rather than write the files locally and
wait until all producers are finished to start the merging process,
the merging process is started as soon as the producers have created a
medium size chunks of data (dozens of megabytes).  Once the merging
process has started, rather than writing the data to local disk, the
data is send directly to the server to be written into the final file and
then to disk, avoiding many unnecessary reads.  Longer term
solutions include upgrading the infrastructure to support the seamless
writing from multiple producing stream within the same process into a
single file. Solutions will be developed in close
collaboration with the main users in order to ease their transition to
the many-cores era.

\subsection{Conclusions}

ROOT is at the core of HEP software as well many other science and
industry frameworks.  Its migration into the many-core era is an
essential stepping stone to enable the eco-system that has been built
upon its libraries to flourish and benefit from the newest hardware
generations.  Many of its features, including a unique customizable
framework to efficiently store and retrieve extremely large data back
and forth between persistent storage and live representation in C++
while supporting significant schema evolution, are fundamental to many
currently running and upcoming experiments and any improvements in
ROOT libraries will greatly benefit them.

Upgrading ROOT infrastructure is a challenging balancing act of
enabling new technologies while supporting existing use patterns to
avoid disruption to running experiments.  However any investment and
progress towards using, enabling and promulgating modern, multitasking
coding patterns will flow through all software layers involved in the
experiments.  Both framework developers and individual researchers
will benefit from the examples and leads that ROOT can provide.

%------------------------------------------------------------------------------
% Philippe Canal and Soon Yung Jun
%------------------------------------------------------------------------------
\section{Geant4 R\&D}

\subsection{Introduction}

Precise modeling of particle interactions with matter in systems with
complex geometries is essential in experimental high-energy physics
(HEP) and certain accelerator and astrophysics applications. Because
of the impact of this type of simulation, it is imperative that the
HEP community has access to computational tools that provide
state-of-the art physical and numerical algorithms and run efficiently
on modern computing hardware. Geant4~\cite{ref:Geant4} is a toolkit for 
the simulation
of particle-matter interactions that has been developed for almost
twenty years by an international collaboration of physicists and
computer scientists. The most commonly used simulation tool in
experimental HEP, Geant4 incorporates physics knowledge from modern HEP and
nuclear physics experiments and theory.

The computing landscape continues to evolve as technologies
improve. Modern-day computers used to support High Energy Physics
(HEP) experiments now have upwards of 32 processors on a single chip,
and that trend will only continue. Furthermore, new generations of
computing hardware such as Graphics Processing Units (GPUs) are
emerging for markets like the enhancement of end-user experience in
gaming and entertainment. Their highly parallel structure makes them
often more effective than general-purpose central processing units
(CPUs) for compute-bound algorithms, where processing of data is done
in highly parallel manner. This technology has great potential for the
high-energy physics community.

The current implementation of Geant4 contains many features that hinder
our ability to make use of modern parallel architectures since Geant4
relies heavily on the object-oriented features of C++ for developing
class hierarchies. Of particular significance is the use of global
resources and the complexity they introduce into the management of
program and processing state. Such an arrangement makes parallelism
through threading difficult. Other contributing factors are its heavy
reliance on C++ inheritance, abstract interfaces, and organization of
of data structures.

The diversity among these forthcoming machines presents a number of
challenges to porting scientific software such as Geant4 and 
achieving good performance.  Extrapolating to the situation over
the next several years, we
anticipate more cores per chip, and these will likely be a
heterogeneous mix of processors, with a few optimized to maximize
single thread throughput, while most are designed to maximize energy
efficiency with wide SIMD data paths. The Advanced Micro Devices (AMD)
Fusion family of Accelerated Processing Units (APU), blending Opteron
CPUs and Radeon GPUs, is just the beginning. This trend will not only
exacerbate performance optimization challenges,but also
simultaneously promote the issues of energy consumption and resilience
to the forefront.  Moreover, as new memory technologies (e.g., phase
change, resistive, spin-transfer torque) begin to appear,
computational scientists will need to learn to exploit the resultant
asymmetric read/write bandwidths and latencies.

\subsection{Geant4 Research and Development for the Future}

To address both the scientific and the computing challenges facing
Geant4, it will be essential to bring together expertise in physics
simulation and modeling with specialists in software systems
engineering, performance analysis and tuning, and algorithm analysis
and development. The overall goal is to move Geant4 into the era of
multi-core and many-core computing, effectively utilizing existing and
future large-scale heterogeneous high-performance computing resources
for both core science simulation data-set generation and also data-set
storage, retrieval, and end-user validation and analysis.

Significant exploration and study is needed in order to move the
Geant4 toolkit toward the long term goal of being able to utilize
large-scale, heterogeneous computing systems, as it was designed for
much simpler platforms. Thus the need for complete studies and work
towards a re-engineering of the underlying system software framework
that retains connections to crucial existing applications and physics
libraries while permitting migration and transition to a new improved
construction.  An extensive reorganization of the underlying
coordination framework will allow for more diverse processing options
(e.g., GPUs) and allow for greater flexibility as computing resources
evolve.

While exploring concurrency, part of the focus should be on
abstracting the overall structure of the Geant4-based codes in terms
of computational flows, algorithms and models. This analysis can be
used to enable the development of code generation schemes and
execution strategies targeting future high-performance architectures.

To enable the use of multiple concurrent and heterogeneous execution
units, Geant4 needs to be re-engineered to have at its core a
concurrent particle propagation engine. Integration of the equations
of motion and determination of particle interaction has very different
structures and computational requirements. Thus a smart dispatcher is
required that can dynamically adapt to its run-time environment,
scheduling bundles of work for resources best suited for the operation
to be performed. The coordination framework must also permit
asynchronous interactions amongst the internal components and with the
surrounding application software infrastructure necessary for event
processing. Such an arrangement allows for more rare and complex
particle interactions to not interfere with more common, less
resource-intensive ones.

\begin{figure}[hbtp]
 \begin{center}
 \resizebox{14cm}{!}{\includegraphics{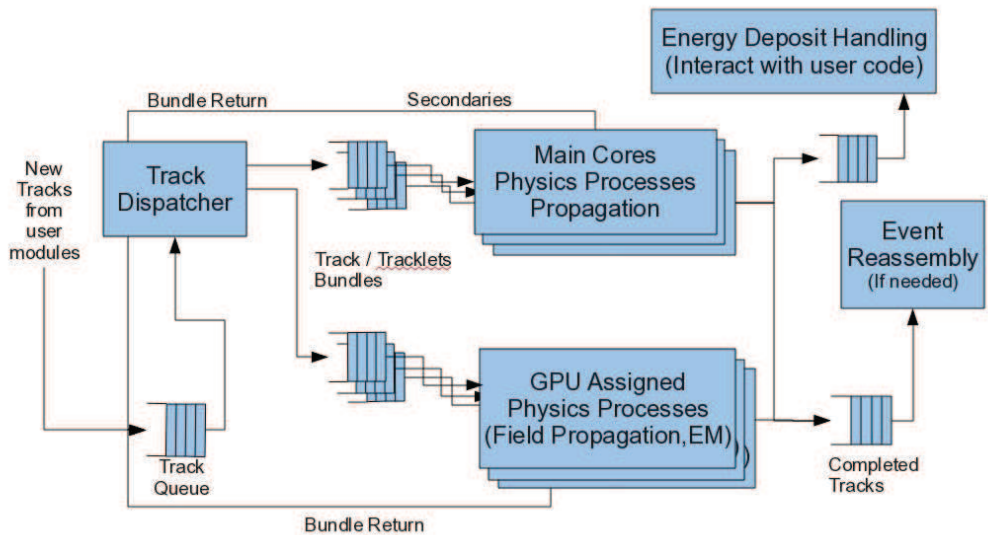}}
 \caption{Simplified view of Geant4 re-engineered processing entities.}
 \label{fig:PicGeant4v4a.eps}
 \end{center}
\end{figure}

Figure~\ref{fig:PicGeant4v4a.eps} depicts a possible structure for
such a concurrent particle propagation engine. A track bundle is a set
of tracks that can be conveniently and efficiently scheduled to go
through the same set of processes which could potentially include
propagation. The role of the track dispatcher is to assemble the
tracks in bundles in such a way as to maximize the efficiency of the
processes that run on them. The dispatcher also selects which
processor or group of processors or cores, for example a GPU enhanced
PC cluster, to schedule the execution of the processes for a specific
track bundle. In order to leverage the on-chip and on-core caches of
general purpose CPUs, the dispatcher might decide to associate a set
of geometrical elements with a specific processor/core and to send to
this processor/core only the bundle of tracks that are traversing
these geometrical elements. It might also gather together tracks that
need to go through processes that have been optimized for GPUs and
send them to one of the graphics cards. Once the tracks are done going
through the simulation, they are added to another stack, which can
also be handled in parallel, to run processes that require all the
finished tracks for a given simulation event, for example to digitize
the energy deposit on the detector elements.

Real-world Geant4 program traces can be used to drive prototype
propagation components. A fundamental change in strategy will be
required, removing the event-by-event and particle-by-particle
processing ensconced in the current framework.  A new strategy will
remove these boundaries and allow large particle vectors to be
assembled across event boundaries where this would enhance processing
efficiency on the target hardware architecture.

A key aspect of this work is adherence to science constraints. Two
important ones are validation of the physics and
reproducibility. Validation tells us that the modules we are working
with give answers that are in statistical agreement with experimental
measurements, or at least in agreement with answers from the
previously validated code. Reproducibility means that results can be
regenerated within a tolerance defined by the scientific community of
users a bit-for-bit agreement where possible. Both validation and
reproducibility are likely to be affected by parallelism and
non-deterministic processing characteristics. Reproducibility will
need to be precisely defined by working with the physics
collaborations. One area of difficulty will be in random-number stream
management. A strategy could be defined to prevent bottlenecks where
random numbers are used and to record necessary seeds to allow
replication of results. Deviations in results, including measurement
uncertainties should be studied.

Geant4 would also benefit from research required to optimize I/O in
heavily multi-threaded systems. In particular, it is important to
understand what data-organization tasks can be achieved by the
simulation itself and what tasks would be more efficiently performed by
separate IO run-time services focused on data collection and
organization.

\subsection{Conclusions}

Geant4 is an important software tool for the HEP community as well as
many other communities. As experiments evolve, the demands on both the
amount of simulated data as well as precision of the simulations will
grow. Geant4 must evolve in order to both take advantage of the
emerging computing technologies and to be able to meet the future
simulation demands in a cost effective way.  Furthermore, this
evolution of Geant4 will enable HEP experimental scientists to take
advantage of current leadership class computing that is available but
not utilized by this community.

Transforming Geant4 will require research to overcome a broad range of
challenges of which categories include systematically improving the
sequential performance of Geant4; re-factoring the code for emerging,
highly parallel computing systems; improving data access, management,
and analysis; exploring novel new programming abstractions, and
reducing the human effort required to handle the upcoming,
Exabyte-scale, globally distributed, data sets generated by Geant4.

There appear to be opportunities for near-term performance enhancement
on individual processors, as well as more profound, long-term changes
that would lead to a new Geant4 that effectively exploits a variety of
future multi-core systems. Taking Geant4 to the next level in terms of
parallelization is paramount for HEP scientists. The next generation
of experiments will demand it and the current ones would profit
greatly.

%------------------------------------------------------------------------------

%------------------------------------------------------------------------------

\end{document}